\documentclass[12pt, letterpaper]{article}
\usepackage[top=1in, bottom=1.5in, left=1in, right=1in]{geometry}
\usepackage[utf8]{inputenc}
\usepackage{booktabs}
\usepackage{hyperref}
\usepackage{graphicx}

\title{Young Stars and their Variability with LSST}
\author{Rosaria (Sara) Bonito and Patrick Hartigan et al. (co-Is in Sect. 1)}
\date{November 2018}

\begin{document}

\maketitle

\begin{abstract}
Young stars exhibit short-term photometric variability caused by 
mass accretion events from circumstellar disks, the presence of dusty
warps within the inner disks, starspots that rotate across the stellar
surfaces, and flares. Long-term variability also occurs owing to
starspot longevity and cycles, and from changes in stellar angular momenta
and activity as the stars age.
We propose to observe the Carina star-forming region in different bands with a
cadence of 30 minutes every night for one week per year to clarify the nature of
both the short-term and long-term variability of the thousands of
young stars in this region. 
By obtaining well-sampled multicolor lightcurves of this dense young cluster,
LSST would acquire the first statistically significant data on how these objects
vary on both short and long timescales. This information will allow us to
relate the observed variability to stellar 
properties such as mass, age, binarity, and to environmental properties
such as location within or exterior to the H~II region,
and to the presence or absence of a circumstellar disk.

\end{abstract}

\section{White Paper Information}

co-PIs: 

Rosaria (Sara) Bonito (rosaria.bonito@inaf.it), INAF - Osservatorio Astronomico di Palermo, Italy 

Patrick Hartigan (hartigan@rice.edu), Physics and Astronomy Dept. Rice University 6100 S. Main Houston, Texas

co-Is:

Laura Venuti, Cornell Center for Astrophysics and Planetary Science (USA)

Mario Guarcello, INAF - Osservatorio Astronomico di Palermo, Italy

Loredana Prisinzano, INAF - Osservatorio Astronomico di Palermo, Italy 

Costanza Argiroffi, Universita' di Palermo, Italy - Dipartimento di Fisica e Chimica

Sergio Messina, INAF - Osservatorio Astrofisico di Catania, Italy

Christopher Johns-Krull (cmj@rice.edu), Department of Physics and Astronomy - MS 108 Rice University

Eric Feigelson, Pennsylvania State University

John Stauffer (stauffer@ipac.caltech.edu), IPAC, California Institute of Technology

Teresa Giannini, INAF - Osservatorio Astronomico di Roma, Italy

Simone Antoniucci, INAF - Osservatorio Astronomico di Roma, Italy

Salvo Sciortino, INAF - Osservatorio Astronomico di Palermo, Italy

Giusi Micela, INAF - Osservatorio Astronomico di Palermo, Italy

Ignazio Pillitteri, INAF - Osservatorio Astronomico di Palermo, Italy

Davide Fedele (davide.fedele@inaf.it), INAF, Osservatorio astrofisico di 
Arcetri, Italy 

Linda Podio (lpodio@arcetri.astro.it), INAF - Osservatorio Astrofisico di Arcetri, Italy

Francesco Damiani, INAF - Osservatorio Astronomico di Palermo, Italy

Peregrine McGehee (peregrine.mcgehee@gmail.com), College of the Canyons, Santa Clarita, California

Rachel Street (rstreet@lco.global), Las Cumbres Observatory

John Gizis (gizis@udel.edu), Department of Physics and Astronomy, University of Delaware

Germano Sacco, INAF - Osservatorio Astrofisico di Arcetri, Italy

Laura Magrini, INAF - Osservatorio Astrofisico di Arcetri, Italy

Ettore Flaccomio, INAF - Osservatorio Astronomico di Palermo, Italy

Salvatore Orlando, INAF - Osservatorio Astronomico di Palermo, Italy

Marco Miceli, Universita' di Palermo, Italy - Dipartimento di Fisica e Chimica

Beate Stelzer (beate.stelzer@astro.uni-tuebingen.de), University of Tuebigen

Julien Fuchs (julien.fuchs@polytechnique.edu), Institute of Applied Physics, Russia; LULI (Laboratoire pour l'Utilisation des Lasers Intenses) - CNRS, Ecole Polytechnique; Commissariat a l'Energie Atomique et aux Energies Alternatives (CEA), Universite Paris-Saclay; Sorbonne Universitss, Universite Pierre et Marie Curie (UPMC) Paris 06, F-91128 Palaiseau cedex, France

Sophia Chen, Institute of Applied Physics, Russia; LULI (Laboratoire pour l'Utilisation des Lasers Intenses) - CNRS, Ecole Polytechnique; Commissariat a l'Energie Atomique et aux Energies Alternatives (CEA), Universite Paris-Saclay; Sorbonne Universitss, Universite Pierre et Marie Curie (UPMC) Paris 06, F-91128 Palaiseau cedex, France

Sergey Pikuz, National Research Nuclear University MEPhI (Moscow Engineering Physics Institute), Moscow 115409, Russia; Joint Institute for High Temperatures, RAS (Russian Academy of Sciences), Moscow 125412, Russia.

Antonio Frasca (antonio.frasca@inaf.it), INAF - Osservatorio Astrofisico di Catania, Italy

Katia Biazzo (katia.biazzo@inaf.it), INAF - Catania Astrophysical Observatory, Italy

Claudio Codella (codella@arcetri.astro.it), INAF - Osservatorio Astrofisico di Arcetri, Italy

Andrea Pastorello (andrea.pastorello@inaf.it), INAF-Oseervatorio Astronomico di Padova, Italy

Juan Manuel Alcala' (juan.alcala@inaf.it), INAF-Osservatorio Astronomico di Capodimonte, Italy

Elvira Covino (elvira.covino@inaf.it), INAF-Osservatorio Astronomico di Capodimonte, Italy

Eleonora Bianchi (eleonora.bianchi@univ-grenoble-alpes.fr),  Univ. Grenoble Alpes, IPAG, 38000 Grenoble, France

Brunella Nisini (brunella.nisini@inaf.it), INAF - Osservatorio Astronomico di Roma, Italy

\vspace{.3in}

with the support of the LSST Transient and Variable Stars Collaboration

\clearpage

\begin{enumerate} 
\item {\bf Science Category:} 
Exploring the transient optical sky

Mapping the Milky Way
\item {\bf Survey Type Category:} Deep Drilling field 
\item {\bf Observing Strategy Category:} 
    \begin{itemize} 
      \item an integrated program with science that hinges on the combination of pointing and detailed observing strategy
    \end{itemize}  
\end{enumerate}  

\clearpage

\section{Scientific Motivation}

Young stellar objects (YSOs) are characterized by photometric variability caused by
several distinct physical processes: mass accretion events from circumstellar disks,
presence of warps in envelopes and disks, the creation of new knots in stellar jets,
stellar rotation, starspots, magnetic cycles, and flares. 
We can study all these phenomena if we acquire both short-term and long-term
lightcurves of a statistically significant sample of YSOs. This white paper outlines
a strategy for maximizing the scientific contribution of LSST to this field of research.

As described below, measuring reliable periods for young stars requires a block
of time where a star-forming region is observed regularly throughout the night
for at least a week, and such observations are also ideal for studying flares,
accretion pulses and disk warps.  The nominal LSST universal cadence complements our
proposed observations by quantifying long-term secular variability. 
Our proposed plan will address open questions related to accretion processes, intrinsic
variability (including eruptive bursts, such as those of EXors-type variables), the evolution
stellar rotation, the spatial distribution of stellar rotation in young clusters,
stellar magnetic activity, and inner disk geometries.

LSST will make a significant impact on our knowledge of young stars
by monitoring the brightnesses and colors of a large sample of these objects. 
Large samples are needed to quantify how the various physical processes listed above
depend upon stellar properties (such as mass, age, binarity),
environmental conditions (stellar location, presence of a circumstellar disks)
and the evolutionary stages of the stars.  This information 
will have a strong impact on the theoretical description of these complex objects.
Knowing the rotation periods of thousands of young stars
and the level and properties of their photometric variability within a given star
forming region (SFR) will be a unique contribution that LSST will provide to this
field of research.  The information derived from LSST photometric time series data will
allow us to learn how angular momentum is distributed among newborn stars, whether
it changes with mass, multiplicity, and location in the cloud, and how it varies
as the stars age.  LSST will allow us to survey an outstanding collection of SFRs
in the Southern hemisphere, including the closest low-mass SFRs (rho Oph, CrA,
Cha I and Lupus), and the most famous intermediate-mass (Orion, Vela) and massive (Carina) SFRs.
Monitoring mass accretion with a survey based on large samples of young stars
is particularly desirable, as accretion is a crucial process in early stellar
evolution. Accretion regulates the star-disk interaction and influences how 
the central object and the protoplanetary disk evolve. 

Photometric variability, on
short- (hours), mid- (days, months), and long-term (years) timescales, is part of
the definition of classical T Tauri stars (CTTSs; Joy 1945).  Amplitude variability
in CTTSs can be up to a few mag (Venuti et al. 2015; Stauffer et al. 2014).
Buster-type objects identified in NGC 2264 (Stauffer et al. 2014) and TW Hya
(Siwak et al. 2018) provide examples of variable accretion, as such short timescale
variability is not produced by long-lived spots or flares.
Different sources of variability (e.g. stellar flares, accretion bursts, absorption
due to warped disks, rotational modulation due to spots) can be easily discriminated
among themselves from their significantly differing observational characteristics (see
examples in NGC 2264, Stauffer et al. 2014, Flaccomio et al. 2018).

Accretion mechanisms in YSOs are easiest to study at short wavelengths 
because the infalling material is heated to several $10^3$ K above the
photospheric temperature as it impacts the stellar surface at near
free-fall velocities. The resulting UV excess luminosity from
the accretion shock is proportional to the total accretion luminosity
and provides a measure of the mass accretion rates (see Gullbring et al. 1998 and Venuti
et al. 2014 for a complete description).  Variations in the accretion rate
are clearly visible as amplitude changes in the blue bands (g filter),
but can still be important even in the z-band for some objects (e.g. TW Hya; Siwak et al. 2018). 
Accretion process in young stars have also been investigated with detailed 3D
magnetohydrodynamic (MHD) models of the infalling material, where such models
account for the observed variability in the inverse P-Cygni line profiles as we
view accretion streams along the line of sight to the star
(Kurosawa \& Romanova 2013; see also the role of
the local absorption for other wavelength emission in Bonito et al. 2014 and
Revet et al. 2017). 

\begin{figure}[h!]
 \centering
 \includegraphics[width=11cm]{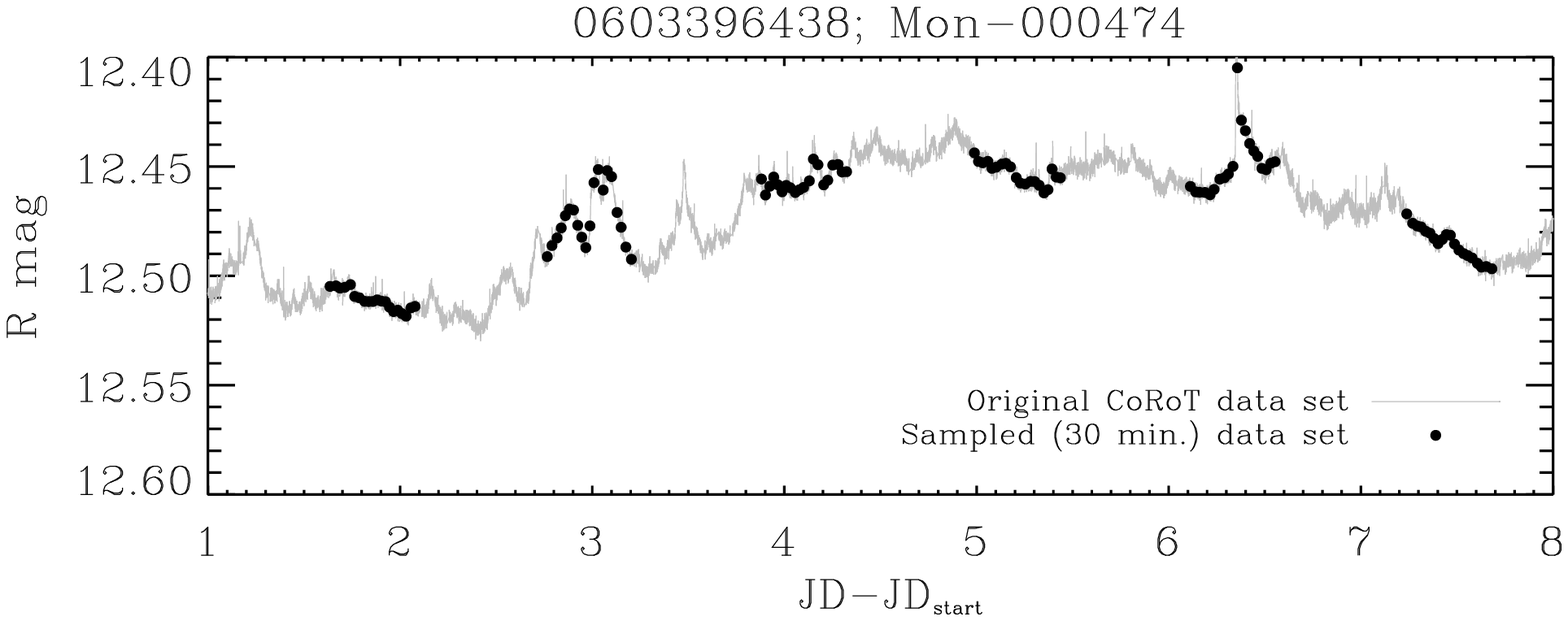}
\includegraphics[width=11cm]{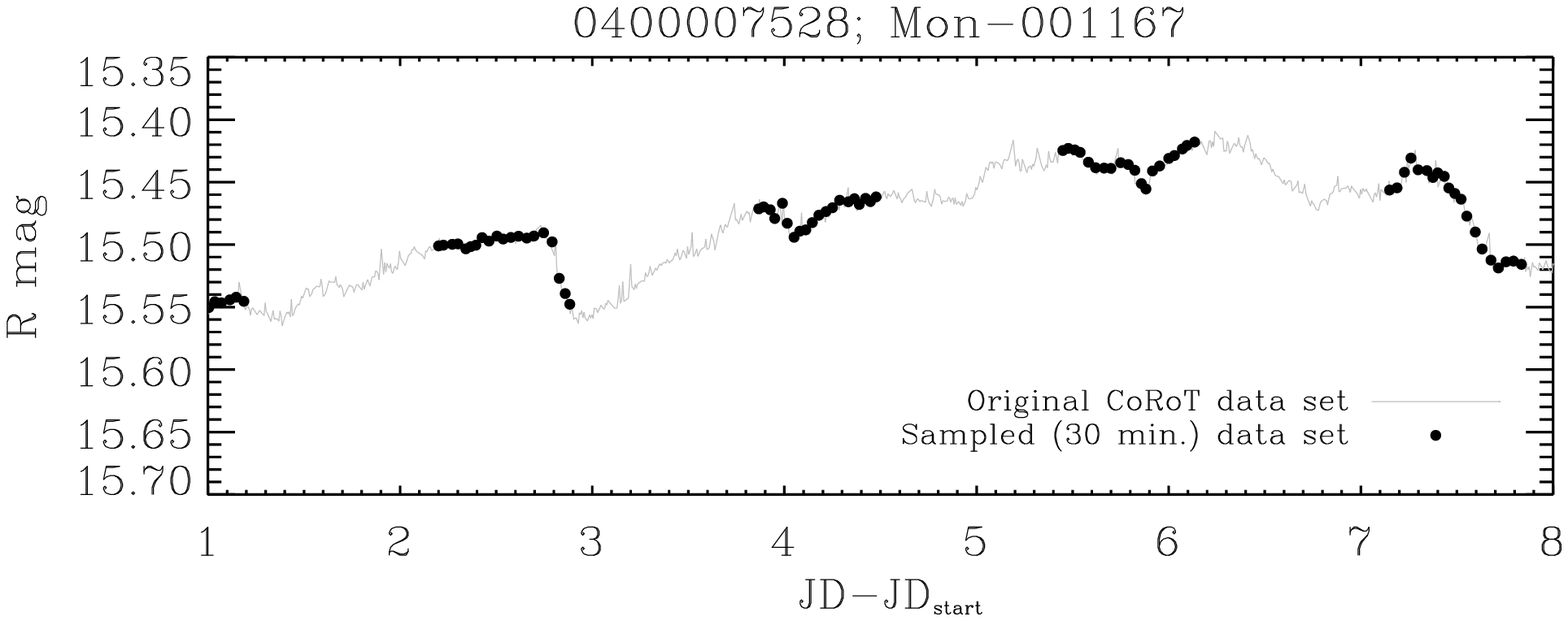}
\includegraphics[width=11cm]{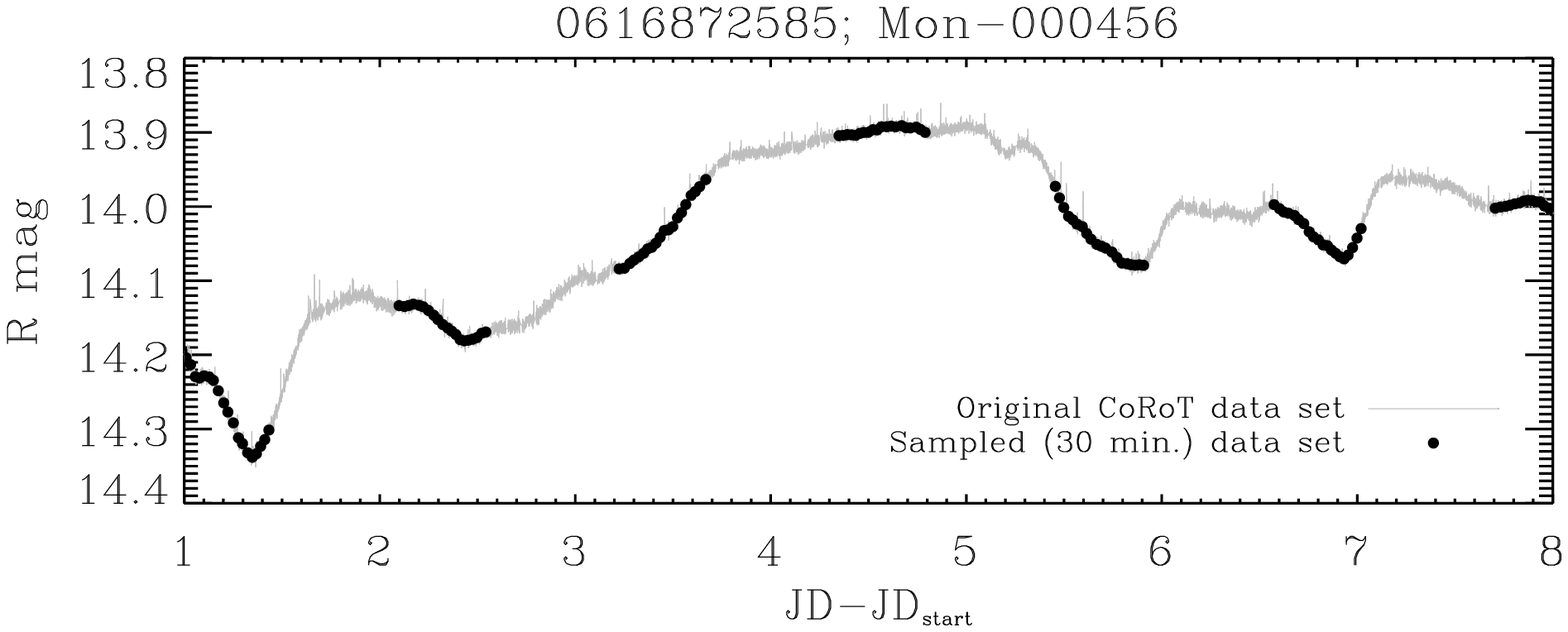}
\includegraphics[width=11cm]{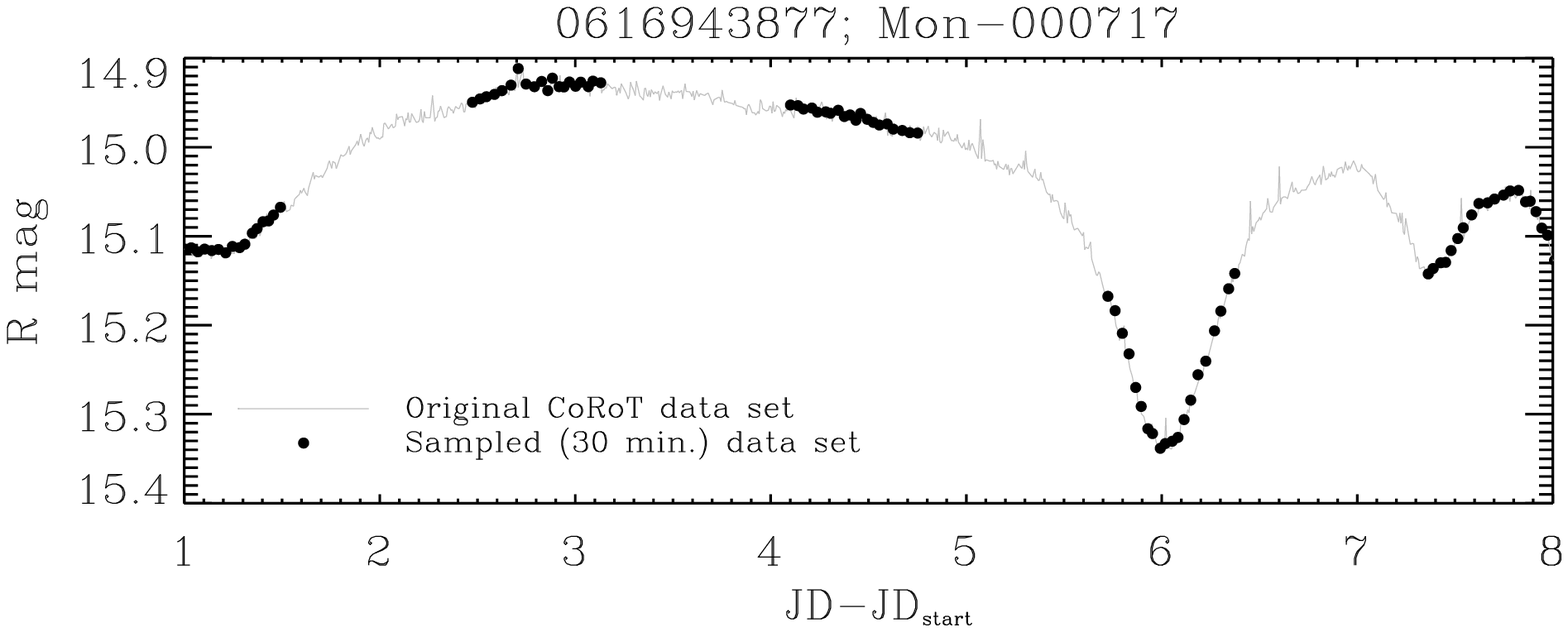}
\caption{\tiny{The light curves in the r-band of young stars showing short-term
non-periodic variability due to accretion, flares, and dips due to a warped disk, as observed with
Corot (see also Stauffer et al. 2014). Bold-faced points mark periods of 
10 hours per night for 7 days with the 30 minutes cadence here requested. Accretion bursts as well
as flares can be discriminated from the LC shape analysis (e.g. in upper panel).
Even with uncertainties per point of $\sim$ 0.02 mag and with 2/3 missing, we still
will be able to reconstruct the proper LC shapes
with the proposed 30 min cadence in each filter.}}
\label{LC}
 \end{figure}

The proposed LSST observations will characterize different class of LCs (see Fig. \ref{LC}),
including light curves dominated by accretion bursts (Stauffer et al. 2014), and light
curves showing periodic or quasi-periodic flux dips (associated to rotating inner
disk warp occulting the stellar emission, Bouvier et al. 2007a; Alencar et al. 2010).
We plan to take advantage of data collected in existing surveys and with previous programs (as
many team members are involved in Gaia-ESO Survey, Chandra, etc.) to characterize
the interesting fields and objects also using a multi-wavelength approach.  Because the
proposed campaigns are only a week long, we plan to ask for spectroscopic data
for the entire duration of the high-cadence observing. Contemporaneous spectroscopy, e.g. with
FLAMES, will be important for detailed comparison with models for individual sources.
A large accretion event is also a powerful way to
select interesting systems for further observations with other instruments. For example, after
an accretion event one could look for evidence of a newly-created jet knot.
We will use accretion events to trigger an alarm to observe the same objects
with other instruments and in different bands (from X-rays to IR). 

The plan is to target one major star-forming region every year. Retargeting the same
region in a subsequent year has the potential to uncover period changes that may arise
from stellar differential rotation, and amplitude variations indicative of stellar cycles. 
Our initial choice of Carina Nebula is well-placed for observations from Chile with LSST,
and guarantees a large number of sources (11,000 members identified; Townsley et al. 2011).
Carina hosts several very massive stars clustered in some of its SFRs. Thus, it contains
regions where disks are affected by photoevaporation and
close encounters (e.g. Guarcello et al. 2016), and quieter regions where they can
evolve unperturbed. We will be able to test the external feedback on disks
evolution from variability-based diagnostics for the first time.  Feigelson et al.
2011 also find thousands of X-ray sources dispersed outside of compact clusters or
clouds, so it is quite reasonable that there will be a vast population beyond
the central 1 sq.deg.  LSST with its unprecedented sensitivity, spatial coverage,
and observing cadence will allow us to employ a statistical approach for the
first time to the comprehension of star formation processes.

\newpage

\newpage

\vspace{.6in}

\section{Technical Description}

\subsection{High-level description}

While sparse coverage of one observation every few days like the
Wide-Fast-Deep (WFD) survey is adequate for identifying sudden changes caused
by large accretion events such as FU Ori and EX Ori outbursts, 
a dedicated campaign to observe star-forming regions at more frequent
time intervals is required to capture the rapidly-rotating periodic stars,
which have periods less than about a day.  Denser phase coverage 
is also needed to follow short-term variations that characterize
accreting and flaring systems.  Embedded and accreting young stars
also undergo significant and rapid color changes owing to
both accretion processes and extinction variations, so it is important to
include multiple filters in any dense coverage campaign.  The goals can be
accomplished by having a week of observations every year where one selected field
is observed once every 30 minutes in g, r, and i bands. A young star with a 2-day
period sampled every 30 minutes provides a data point every 0.01 in phase. For
the best-case scenario, observing for 7 nights and 10 hours per night would yield
140 photometric points in each filter.
Depending on the period aliasing, this coverage should populate the phases well
enough to identify most of the large starspots on the stellar photospheres.

At the beginning of LSST operations we argue that a targeted test field (Carina
Nebula) should be observed in the above manner to illustrate what can be done with
LSST in this mode. If successful, in
subsequent years we will either choose a different region or possibly return
to the same regions to monitor slow changes in periods or amplitudes that
may arise from differential rotation or starspot cycles.
Combining a densely-packed short-interval dataset with a
sparse but long baseline study maximizes the scientific return for both methods,
and allows LSST to address all of the accretion and rotational variability
associated with young stars.  

\vspace{.3in}

\subsection{Footprint -- pointings, regions and/or constraints}

The first target requested for the first year of observations in this program is
the Carina Nebula.  The pointing should be centered on Eta Carinae
(RA 10:45:03.53, Dec -59:41:04.1, ICRS coord., ep=J2000).
Other SFRs can be considered in the future once the data
analysis of first year observations is obtained in the context of this program.
Examples of other SFRs visible from Chile include Orion Nebular Cluster, NGC 2264,
NGC 6530, and NGC 6611. 

Galactic star formation regions are largely found at low Galactic latitudes or
within the Gould Belt.  As such, study of young stars with LSST is closely
tied to other science goals concerning the Milky Way Disk and is subject to the
concerns of both crowded field photometry and the observing cadence along the Milky
Way. However, DECam observations at the CTIO 4-m that reached depths similar to
those proposed for LSST show negligible crowding in the optical, and $< 5\%$ crowding
at z in Carina. In this case, extinction in the molecular clouds helps by
significantly lowering the frequency of background contamination.
Owing to extinction in the dark clouds, source confusion will generally not be an
issue (as evidenced by typical deep optical images of such regions).
Most young stars congregate into clusters in specific regions, though there is an
older population that is more distributed. The vast majority are within about
25 degrees of the galactic plane.
By observing young clusters we take advantage of the ability to image a large number of
coeval pre-main-sequence stars simultaneously. Field stars will be useful as controls
to ensure than any bursts in the LCs of young stars are real (e.g. Stauffer et al. 2014).

\subsection{Image quality}

Better seeing helps to resolve close binaries, and is also important
for regions where contamination comes into play, for example, in the plane but away from dark clouds.

\subsection{Individual image depth and/or sky brightness}

Some of the fainter objects will be affected if the Moon is very bright and close.
However generally these constraints are secondary and do 
not affect the design of the survey.  Considering DECam 60-sec r-band images, it
is possible to get very good photometry down through r=20, and it starts to become too
noisy for good light curves once r=21 for data collected on Carina with a $100\%$ full Moon.
According to the CTIO website, the sky brightness is about 18 mag/sq-arcsec for a
full Moon.

\subsection{Co-added image depth and/or total number of visits}

Deep coadded frames are a secondary priority.

\subsection{Number of visits within a night}

Accretion variability in classical T Tauri stars typically ranges from 30 min to several hours
or longer (e.g. TW Hya, Siwak et al. 2018).  Sampling on timescales short enough to trace
the short-term variations ($<$hours; flare, burst) but extending over timescales
relevant to various processes is needed to achieve a detailed physical description
of the mechanisms at work at the stellar surface and in the star-disk interface (Venuti et al. 2015).
In general, a small number of epochs and irregular cadence does not allow one to
discriminate the physical nature of the variability (see discussion on this issue
in Stauffer et al. 2014).  We therefore request observations once every 30 minutes,
10 hours per night, for 7 nights. This would yield 20 photometric points in each
filter (each night), and will populate the phases well enough to identify most
of the large starspots on the stellar photospheres and track other expected
sources of variability.  Data in each band (changing every 30
minutes g, r, and i filters) gives its own lightcurve, making it possible to
follow how the colors vary with phase.

Siwak et al. 2018 derived a rate of 0.59 flare/day or 0.94
(flare or accretion burst)/day for TW~Hya, with accretion bursts lasting up
to 30 min.  Therefore, we expect to observe several peaks in the LCs collected
each night with the proposed cadence. Models suggest accretion cooling
timescales of 30 min to several hours in accordance with observations of the shortest bursts in
BP Tau (0.6 h; see discussion in Siwak et al. 2018).

\subsection{Distribution of visits over time}

In this white paper we ask for an annual week-long run of 10 hours per night
of a targeted region, with observations taken every 30 min to complement the universal cadence of one
observation every few days.
Variability due to stellar activity, accretion process including eruptive bursts
(EXors), rotation, etc., will all benefit from higher cadence observations exploring
clusters with different ages, metallicity, and location.

As an example,
Venuti et al. 2014 explored the mid-term variability in NGC 2264 and measured the
UV excess (and correspondingly computed the mass accretion rate) from each observing
epoch during the CFHT r-band (and u-band) monitoring, obtaining $\approx17$ points
distributed over the 2-week long survey to probe variability.
Stauffer et al. 2014
found a burst frequency of 0.2 peak per day and typical total duration of isolated
accretion burst of 1 d. Therefore we expect to be able to observe 1 peak in about
5 days, in good agreement with our proposal to observe with LSST with a rolling
cadence of 1 week every 30 min in g, r, and i filters: the higher cadence requested
would allow the detection of many shorter duration (hours) events.

If part of a night is lost due to poor weather, the science
is still achievable (there is no need to the whole sequence to be redone). However,
if several days are lost then a typical phase coverage will become sparse, so
operations staff should use their best-efforts to schedule the observations
during a week of continuously usable weather.

\textbf{Figure of Merit (FoM)}:
In the proposed program, we expect to collect 140 points in each filter (g, r, and i) in 1 week.
The Wide Fast Deep (WFD) main survey will collect 80, 180, and 180
visits in g, r, and i filters respectively in ten years of LSST Survey.
In the WFD scenario, we would collect 0.15, 0.35, and 0.35 observations
per week in g, r, and i filter respectively as opposed to 140 per week in the proposed
rolling cadence. Recovering periods from WFD data will be impossible if the periods
vary even slightly over a 10-year timespan. For example, a young star with
a rotation period of 2 days will experience 1825 rotations in 10 years. If the period slows
by just one minute over the 10 years time interval, the lightcurve will shift by 0.63 in phase,
making it impossible to properly phase such sparse data.
The planned number of visits in the Galactic Plane in 10 years (p.55 SB, sect.3.1) are especially
low, $<30$ in all filters ($< 1$ every four months), and completely unsuited for
period studies.

\vspace{.6in}

\subsection{Filter choice}

We request a rotation of g, r, and i filters every 30 minutes for 10 hours each
night for 1 week each year (possibly changing the pointing every year, but
starting with Carina Nebula the first year).  Typical T~Tauri stars will be
too faint to observe at u using a rapid cadence with LSST, but the g-band is
blue enough to use as an accretion diagnostic.  In CTTSs, blue band fluxes rise more
strongly during accretion events, so we can distinguish accretion from extinction events
if red magnitudes are also available.  Furthermore, the r-band data will be
important as it allows us to construct a color magnitude-diagram r vs. g-r.
We should be able to identify the WTTS members of the cluster, while a bluer spread characterizes the CTTS members, as the  blue band excess is related to the accretion activity only present in CTTSs. Therefore, an important discrimination between WTTS and CTTS among the cluster members can be performed. 

CTTSs show higher levels of
variability, both in the optical (r band) and, more markedly in the blue bands
such as g, with typical photometric amplitudes about three times those measured
for WTTSs (see Venuti et al. 2015 for the u-band in particular;
see Siwak et al. 2018 for the g band).  The LCs which are burst dominated, and therefore
related to accretion process, have been found to be $55 - 80\%$ of the YSO with
the strongest UV excesses in NGC 2264 (Stauffer et al. 2014).
Flares also occur in WTTS as a consequence of high chromospheric activity.  Flaring in WTTS
can also be monitored, though the rapid decline of chromospheric flares requires a rapid 
cadence to capture correctly.  The blue band allows us to follow mass accretion
variability, while r (for most objects) will be dominated by photospheric flux. 

More colors are always useful, but having a photospheric color index (r-i) plus one
accretion measure (g-r) are the science drivers for filter choices.
Variability in different colours helps to
discriminate between hot spots, cold spots, and circumstellar extinction. 
A color-color diagram also helps to correct erroneously classified
non-accreting members. In fact, the accretion process is intrinsically variable
and during previous surveys low accretors could be misinterpreted as a non-accretor
members. 

Classifications based on color can be confirmed spectroscopically with, for example, FLAMES
observations of the H$\alpha$ emission line (see also Bonito et al.
2013 and Bonito et al. in preparation).  In the g vs
g-i diagram, we can compare the location of CTTSs with accretion burst, dips in the
LCs, and WTTSs (see the example of NGC 2264 members in Stauffer et al. 2014, Fig. 9). 
The three different groups of stars appear coeval in general, with a smaller
dispersion for the non-accreting component (see also Lamm et al. 2004).  Stars with
LCs dominated by accretion bursts have a location on the g vs g-i diagram affected by
hot spots formed by accretion (Stauffer et al. 2014).

\subsection{Exposure constraints}

A rotation of 30 seconds exposuree in g, r, and i filters every 30 minutes for 10 hours per night for 7 consecutive days each year

\subsection{Other constraints}
None

\subsection{Estimated time requirement}

We request a week of observations every year once every 30 minutes in g, r, and i, bands. 
Observing for 7 consecutive nights and 10 hours per night would yield 140 photometric points in each filter.
The total overhead is 34 seconds per visit as shown below. During the week that the campaign is
active, it will use approximately 25\%\ of the available observing time on those nights.

\begin{enumerate}
\item g-band: 120 sec, for slew and setting; 30 sec exposure; 2 sec shutter; 2 sec readout
\item r-band: 120 sec, to change the filter; 30 sec exposure; 2 sec shutter; 2 sec readout
\item i-band: 120 sec, to change the filter; 30 sec exposure; 2 sec readout
\end{enumerate}
 
\vspace{.3in}

\begin{table}[ht]
    \centering
    \begin{tabular}{l|l|l|l}
        \toprule
        Properties & Importance \hspace{.3in} \\
        \midrule
        Image quality & 2    \\
        Sky brightness & 2 \\
        Individual image depth & 2  \\
        Co-added image depth & 2  \\
        Number of exposures in a visit   & 2  \\
        Number of visits (in a night)  & 1  \\ 
        Total number of visits & 1  \\
        Time between visits (in a night) & 1 \\
        Time between visits (between nights)  & 1  \\
        Long-term gaps between visits & 3\\
        Other (please add other constraints as needed) & 3\\
        \bottomrule
    \end{tabular}
    \caption{{\bf Constraint Rankings:} Summary of the relative importance of
various survey strategy constraints. Please rank the importance of each of these
considerations, from 1=very important, 2=somewhat important, 3=not important. If
a given constraint depends on other parameters in the table, but these other
parameters are not important in themselves, please only mark the final constraint
as important. For example, individual image depth depends on image quality, sky
brightness, and number of exposures in a visit; if your science depends on the
individual image depth but not directly on the other parameters, individual image
depth would be `1' and the other parameters could be marked as `3', giving us the
most flexibility when determining the composition of a visit, for example.}
        \label{tab:obs_constraints}
\end{table}

\subsection{Technical trades}
This program is a rolling cadence of observations that should be performed in g,
r, and i filters every 30 minutes each night (10 hours) for 7 consecutive days
each year, to properly reconstruct the LC shapes to discriminate among different
possible physical mechanisms at work in young stars.

\section{Performance Evaluation}

To quantify YSO studies with LSST, we consider V927 Tau, a rather faint,
moderately-reddened 0.2 $M_\odot$ young star in the Taurus cloud as a target goal. 
Scaling the SDSS colors of V927 Tau, a typical low-mass T Tauri star, to Carina
gives u=24.1, g=21.6, r=20.2, i=18.9 and z=17.9 mag.  These numbers motivate 
our choice of filters (i, r, and g) and allow us to say with confidence
that we will get good r and i LCs of Carina's low mass T Tauri stars with LSST.
60-second exposures with DecCAM on the CTIO 4-m 
were able to get to r=20.2 with good precision with a $100\%$ full Moon. 

Note that 30-sigma limits for each 4-minute dither sequence (good lightcurves) are
u=21.6, g=22.5, r=22.3, i=22.0, and z=21.3.  For reference, a typical young star
in the Carina X-ray catalog has an i-magnitude of 18. The
universal cadence option of $2\times15$ sec exposures will yield sigma = 0.02 mag
for r=21.8, a magnitude fainter than V 927 Tau would be in Carina. This photometric
uncertainty suffices to recover a typical period from such an object. LSST will
determine periods to near the hydrogen burning limit with nominal r-band
exposure times for a region like Carina.

\vspace{.6in}

\section{Acknowledgements}

This work was developed within the Transients and Variable Stars Science Collaboration (TVS) and the author acknowledges the support of TVS in the preparation of this paper.

\section{References}

Alencar et al. 2010, A\&A, 519, 88 \\
Bonito et al. 2014, ApJ, 795, 34 \\
Bouvier et al. 2007, A\&A, 463, 1017 \\
Feigelson et al. 2011, ApJS, 194, 9 \\
Flaccomio et al. 2018 \\
Guarcello et al. 2016 \\
Gullbring et al. 1998 ApJ 492, 323 \\
Joy 1945, ApJ, 102, 168J \\
Kurosawa and Romanova 2013, MNRAS, 431, 2673 \\
Revet et al. 2017, Science Advances, 3, 0982 \\
Siwak et al. MNRAS, 478, 758 \\
Stauffer et al. 2014, AJ, 147, 83 \\
Venuti et al. 2014, A\&A, 570, 82 \\
Venuti et al. 2015, A\&A, 581, 66 \\

\end{document}